\documentclass[showpacs,floatfix,superscriptaddress]{revtex4}
\usepackage{graphicx}
\usepackage{caption}
\usepackage{subfig}
\usepackage{epsfig}
\usepackage{bm}
\usepackage[utf8]{inputenc}
\usepackage{amssymb}
\usepackage{float}
\usepackage{amsmath}
\usepackage{dcolumn}
\usepackage{latexsym}
\usepackage{subfig}
\usepackage{amsthm}
\usepackage{framed}
\usepackage{cancel}
\usepackage[colorlinks]{hyperref}
\usepackage[usenames,dvipsnames]{color}
\usepackage{multirow}
\usepackage[pdftex]{pict2e}
\usepackage[dvipsnames]{xcolor}

\hypersetup{
     breaklinks=true,
    pdfstartview={FitH},    
    colorlinks=true,       
    linkcolor=blue,          
    citecolor=red,        
    filecolor=magenta,      
    urlcolor=blue,           
    anchorcolor=green,      
    linktocpage=true
}

\def\doi{http://doi.org}

\begin{document}

\title{Shadow behavior of an EMSG charged black hole}

\author{Fateme Aliyan}
\email[]{f.aliyan03@umz.ac.ir}
\author{Kourosh Nozari}
\email[]{knozari@umz.ac.ir (Corresponding Author)}
\affiliation{Department of Theoretical Physics, Faculty of Basic Sciences, University of Mazandaran,\\
P. O. Box 47416-95447, Babolsar, Iran}

\begin{abstract}
Recent shadow images of Sgr A* and M87* captured by Event Horizon Telescope (EHT) collaboration confirm the
existence of black holes or their possible alternatives in the center of galaxies.
On the other hand the new image of Sgr A* in polarized light suggests a Magnetic field spiraling at the Edge of the Milky Way’s Central Black Hole. Due to gravitational lensing effect, bending of light in the background geometry of the black hole casts a shadow.
In recent years, black holes and their properties have been vastly studied in the framework of General Relativity and
other modified theories of gravity. One of the possibilities to generalize GR is Energy-Momentum Squared
Gravity (EMSG) which is constructed by adding a term proportional to
$\mathbf{T}^{2} =\mathbf{T}^{\alpha\beta}\mathbf{T}_{\alpha\beta}$ (where $\mathbf{T}_{\alpha\beta}$ is the
energy-momentum tensor) in the gravitational action. It is important to mention that EMSG modifies all matter field's equation which leads to add some non-linear terms to Maxwell equations.
 EMSG theory as a modified theory of gravity predicts an asymptotically de Sitter charged black hole
whose shadow cast and other related characteristics have not been examined yet.
Hence we consider the EMSG charged black hole and investigate the shadow shape of this kind of black hole solution in confrontation with EHT results. In the case of non-linear electrodynamics the photon's path is null on some effective metric. by deriving the effective metric of EMSG charged black hole we study the null geodesics of the effective metric in Hamilton-Jacobi method.  we find the photon orbits and compute
the shadow size of this black hole. Then we examine how electric charge and the coupling constant of the
EMSG affect the shadow size of the black hole
in a positively accelerated expanding universe (with a positive cosmological constant).
We explore the viable values of these parameters constrained by EHT data by comparing the shadow
radius of EMSG charged black hole with the shadow size of Sgr A*.
We show for instance that for $Q = 0.1$ in appropriate units, the coupling constant should be in
the range of $0.01\leq \eta \leq 0.02$ in order to EMSG charged black hole to be the Sgr A*.
Consecutively we obtain that in the case of $\eta=0.01$ the range of the electric
charge could be $0.01\leq Q \leq0.3$ in the adopted units. We observe that by enhancing the effect of the
electric charge, the shadow size of this EMSG charged black hole increases accordingly
By treating the energy emission rate of EMSG charged black hole, we demonstrate that for the small amount of the
electric charge and large values of the coupling constant, the black hole evaporates faster.\\

\end{abstract}


\maketitle

 \section{INTRODUCTION}

The acquisition of the near horizon images of the supermassive black holes M87*
\cite{Akiyama2019,Akiyama20191,Akiyama20192,Akiyama20193,Akiyama20194,Akiyama20195} and
Sgr A* \cite{Akiyama20198,Akiyama2019b,Akiyama20199,Akiyama201910}
by the Event Horizon Telescope (EHT) collaboration is more or less direct proof of the existence of
black holes as claimed by General Relativity and other theories of gravity. Based on common belief, black holes
are the key objects to reveal the mystery of the unification of
quantum mechanics and gravity \cite{Barack2019} as long as they deal with the most extreme regions
of spacetime \cite{Schwarzschild1916,Penrose1965,Bambi2018}.
According to the cosmic censorship hypothesis, by gravitational collapse of an object into a state of singularity, the black hole
will be formed so that the singularity will be covered with an event horizon. Since classically neither particle nor radiation
can return from the black hole's event horizon, it is totally unobservable by anyone who is located outside the black hole.
However, as a result of the gravitational lensing phenomena \cite{Perlick:2010zh}, in the presence of a background
light source the observer will find a dark spot in the direction of the black hole which is known as the black hole
shadow. Since the appearance of the initial idea about black holes and other astrophysical objects'
shadows \cite{Synge1966,Luminet1979}, many analytical and numerical studies on the
various aspects of black hole shadow have been reported in literature; see just for
instances~\cite{Perlick2022,Wang2022,Ayzenberg2023} and references therein.

Afterward releasing the images of the supermassive black holes M87* and Sgr A*, the EHT collaboration has uncovered a new view of the black hole in
polarized lights \cite{Akiyama20197,Akiyama20196,Akiyama201911,Akiyama2019c,Akiyama2024,Akiyama20241}. These observation show strong and organized magnetic fields spiraling from the edge of the supermassive black holes M87* and Sgr A*. Since magnetic fields may be common to all black holes, investigation has looked into how these strong magnetic fields influence the shadow images of black holes \cite{Junior2021}. In addition, the study of black hole shadow in nonlinear electrodynamics has been considered \cite{okyay2022,zhong2021,zhong2021a,Allahyari2020}.

On the other hand, the first serious attempt to modify gravity backs to the late 19th century in the way to model
Maxwell’s electrodynamics in modification of Newtonian gravity. From the beginning of GR,
huge attention has been devoted to generalizing it \cite{Eric2014}.
Indeed the initial attempt to modify GR was made by Einstein himself by adding a Cosmological Constant ($\Lambda$)
term to the original version of GR's field equations. At present, observations conducted towards the dark matter and dark
energy, motivated more and more physicists to devote their efforts to generalize GR, see for instance
\cite{Nojiri2007,DeFelice2010,Nojiri2011,Capozziello2011,Nojiri2017}. The Energy-Momentum Squared Gravity (EMSG) is among these attempts towards modification of the GR in the
favor of observations~\cite{Katrc2014,Roshan2016}. EMSG theory is the result of adding
the term $\mathbf{T}^{2} =\mathbf{T}^{\alpha\beta}\mathbf{T}_{\alpha\beta}$ (with $\mathbf{T}_{\alpha\beta}$ as
the energy-momentum tensor) to the Einstein-Hilbert action with a coupling constant $\eta$ whose
value can be constrained by observations. This modified gravity
theory has attracted considerable attention in recent years. Just for some instances, cosmological dynamics with a scalar field through
the Noether Symmetry approach has been considered in the Energy-Momentum Squared Gravity in Ref.~\cite{Sharif:2022ruy}. Cosmological inflation within
the energy-momentum squared gravity in confrontation with recent observational data has been studied in Ref.~\cite{HosseiniMansoori:2023zop}.
The issue of Noether's Symmetries of an anisotropic universe within Energy-Momentum Squared
Gravity has been investigated in Ref.~\cite{Sharif:2021hyj}. A possible realization of the Gödel-type model
universes in the Energy-Momentum-squared gravity is studied in Ref.~\cite{Canuto:2023gdv}. The issue of decoupled Gravastars in
the framework of the Energy-Momentum Squared Gravity is investigated in Ref.~\cite{Sharif:2023uac}.
Recently the cosmological evolution of the spherical over densities in Energy-Momentum-Squared
Gravity is studied in Ref.~\cite{Farsi:2023gsz}. Also bending of light and gravitational lensing are
analyzed in \cite{Nazari:2022fbn} within the energy-momentum-squared gravity.
The Energy-Momentum squared gravity is constrained via binary pulsar observational data in Ref.~\cite{Nazari:2022xhv}.
Possible realization of the cosmological inflation in Energy-momentum Squared Gravity and its observational viability in
confrontation with Planck 2018 data are investigated in Ref.~\cite{Faraji:2021laz}. Seeking for viable Wormhole
solutions in the context of the Energy-Momentum Squared Gravity \cite{Sharif:2021gdv}, Non-minimal extension of the Energy-momentum Squared
Gravity \cite{Shahidi:2021lqf}, Noether's Symmetries of the Energy-Momentum Squared Gravity \cite{Sharif:2020rby},
Constraints on Energy-Momentum Squared Gravity from cosmic chronometers and Supernova type Ia data \cite{Ranjit:2020syg}, thermodynamics of
the apparent horizon in a universe governed by a generalized Energy-Momentum Squared Gravity~\cite{Rudra:2020rhs} and generalized Energy-Momentum Squared
Gravity in the Palatini formalism \cite{Nazari:2020gnu} are some recent studies in this field.
Also quark stars with color-flavor locked property have been studied in Energy-Momentum Squared Gravity by Singh et. al.~\cite{Singh:2020bdv}.
Possible temporal variation of the universal gravitational ``constant" and the speed of light in the
Energy-Momentum Squared Gravity have been investigated in Ref.~\cite{Bhattacharjee:2020fgl}. Jeans analysis~\cite{Kazemi:2020hep},
viability of the bouncing cosmology~\cite{Barbar:2019rfn}, Eikonal black hole ringings~\cite{Chen:2019dip},
dynamical system analysis~\cite{Bahamonde:2019urw}, screening of the $\Lambda$~\cite{Akarsu:2019ygx},
cosmological implications of the scale-independence and pseudo-non-minimal interactions in the dark matter and relativistic relics~\cite{Akarsu:2018aro},
compact stars~\cite{Nari:2018aqs}, constraint from neutron stars, and cosmological implications~\cite{Akarsu:2018zxl} and some new cosmological
models~\cite{Akarsu:2018zxl, Board:2017ign} are among the works devoted to the various aspects of the Energy-Momentum Squared Gravity in recent years.

Among all these works, there is an apparent gap in this field regarding the shadow of an EMSG-charged black hole and its
observational status, which we have decided to fill this gap in this paper. Theoretically, black holes are a
family of solutions of the theory of gravity characterized by three quantities; mass,
spin and electric charge. The spherically symmetric charged black hole solution in EMSG theory has been found
in Ref.~\cite{Roshan2016}. In the context of the electromagnetic field, EMSG introduces some non-linear components to Maxwell’s equations. This makes EMSG similar to Born-Infeld nonlinear electrodynamics \cite{Born1934}. However, the specific field equations in EMSG are different from those in Born-Infeld theory. In nonlinear electrodynamics, photons follow paths in an “effective geometry” instead of actual spacetime \cite{Plebansky1968} which implies that in a complex vacuum, light behaves like electromagnetic waves moving through a medium that alters their motion \cite{Dittrich1998}. The advancements in nonlinear quantum electrodynamics have significantly improved our understanding of light propagation, including the geometry of light propagation \cite{Cai2004} and the deflection of magnet stars in Born-Infeld theory \cite{Kim2022}.
 No study on the shadow of this black hole and its status in confrontation with EHT data has been reported
so far, which made us interested in investigating this phenomenon in the background of the EMSG-charged black hole.
For this purpose, in this paper we considered a positive cosmological constant to be responsible for the
cosmic positively accelerated expansion. Then through the EHT observation, we constrain the coupling constant of
EMSG theory and electric charge. Finally, we address the issue of acceleration bounds in this black hole spacetime background.

This paper is organized as follows. In section \ref{se1} we introduce the EMSG action and corresponding field equations shortly.
Then we express the EMSG-charged black hole solutions in section \ref{se2}. In section \ref{se9} we derive the effective metric for EMSG charged black hole. Section \ref{se3} deals with the investigation of photon
motion in the EMSG-charged black hole spacetime and following that in section \ref{se4} we demonstrate
the geometrical shape of the EMSG charged black hole. Then we constrain the results via EHT data in
section \ref{se5}.
In section \ref{se7} we study the energy emission rate for the black hole.
Finally, section \ref{se8} is devoted to the summary and conclusion.

\section{ACTION AND FIELD EQUATIONS OF EMSG}\label{se1}

The total action of the Energy-Momentum-Squared Gravity includes the following terms~\cite{Katrc2014,Roshan2016}
\begin{equation}\label{eq1}
  \mathcal{S}_{EMSG}=S_{GR}+\mathcal{S}_{\mathbf{T}}+\mathcal{S}_{M}\,,
\end{equation}
in which  $\mathcal{S}_{GR}$ is the Einstein-Hilbert action, the modification term $\mathcal{S}_{\mathbf{T}}$ is the action for
the energy-momentum tensor as the characteristic term of this model and $\mathcal{S}_{M}$ is the matter action of the model.
These terms read as follows
 \begin{equation}\label{eq2}
\mathcal{S}_{GR} = \frac{1}{2\kappa}\int \sqrt{-g}(R-2\Lambda)d^{4}x\,,
\end{equation}
\begin{equation}\label{eq3}
\mathcal{S}_{\mathbf{T}} = \frac{\eta}{2\kappa}\int \sqrt{-g}(\mathbf{T}^{2})d^{4}x\,,
\end{equation}
\begin{equation}\label{eq4}
\mathcal{S}_{M}=\int \sqrt{-g} \mathcal{L}_{M} d^{4}x\,,
\end{equation}
where $g$ is the metric determinant, $\mathcal{L}_{M}$ is the matter Lagrangian density, $\kappa=8\pi G$,
$R$ is the Ricci scalar, $\Lambda$ is the cosmological constant, $\mathbf{T}^{2}=T_{\mu\nu}T^{\mu\nu}$
is the energy-momentum tensor squared and $\eta$ is a coupling constant which can be
positive or negative. By setting $\eta=0$, the EMSG theory reduces to Einstein's gravity.
The EMSG gravitational field equations are obtained by
varying the total action (\ref{eq1}) with respect to the metric tensor $g_{\alpha\beta}$
\begin{equation}\label{eq5}
G_{\alpha\beta}+\Lambda g_{\alpha\beta}=\kappa  T^{(eff)}_{\alpha\beta}\,,
\end{equation}
where $G_{\alpha\beta} = R_{\alpha\beta}-\frac{1}{2}R g_{\alpha\beta}$ is the Einstein tensor. In this setup the
effective energy-momentum tensor is defined as follows
\begin{equation}\label{eq6}
 T^{(eff)}_{\alpha\beta}= T_{\alpha\beta}+ 2 \frac{\eta}{\kappa}({\cal{T}}_{\alpha\beta}+ T^{\lambda}_{\alpha}
 T_{\lambda\beta}-\frac{1}{4}g_{\alpha\beta}\mathbf{T}^{2})\,,
\end{equation}
where
\begin{equation}\label{eq7}
 {\cal{T}}_{\alpha\beta}=T^{\rho\sigma} \frac{\delta T_{\rho\sigma}}{\delta g^{\alpha\beta}}.
\end{equation}
The effective energy-momentum tensor $T^{(eff)}_{\alpha\beta}$ can be determined if the matter action is known.
In other words, the energy-momentum tensor $T_{\alpha\beta}$ must be obtained by varying the
matter action with respect to the gravitational degrees of freedom, $T_{\alpha\beta}=
-\frac{2}{\sqrt{-g}} \delta(\sqrt{-g}\mathcal{L}_{M})/ \delta g^{\alpha\beta}$, then ${\cal{T}}_{\alpha\beta}$ can be written as

\begin{equation}\label{eq8}
{\cal{T}}_{\alpha\beta}= -\mathcal{L}_{M}(T_{\alpha\beta}-T \frac{g_{\alpha\beta}}{2})- \frac{1}{2}T
 T_{\alpha\beta}-2T^{\rho\sigma}\frac{\partial^{2}\mathcal{L}_{M}}{\partial g^{\rho\sigma} g^{\alpha\beta}}\,.
\end{equation}
Therefore, the field equations (\ref{eq5}) are perfectly known for the Lagrangian density of a particular matter distribution.
It is clear that EMSG corresponds to GR in vacuum where the energy density of matter is zero. Therefore,
the Schwarzschild-de Sitter and Kerr metrics are also solutions to the EMSG field equations.
In the presence of an exterior electromagnetic field, the EMSG field equations take a different form than GR as we will see in what follows.

\section{CHARGED BLACK HOLES IN EMSG}\label{se2}

The exact black hole exterior solution in the presence of an electromagnetic field has been derived in Ref.~\cite{Roshan2016}. An electromagnetic field
$\mathcal{F}_{\mu\nu}=\partial_{\mu}A_{\nu}-\partial_{\nu}A_{\mu}$ around a black hole satisfies the following energy-momentum tensor
\begin{equation}\label{eq9}
 T_{\alpha\beta}=\mathcal{F}^{\rho}_{\alpha}\mathcal{F}_{\rho\beta}-\frac{1}{4}g_{\alpha\beta}\mathcal{F}^{2}\,.
\end{equation}
Therefore, the EMSG field equations will be in the following form~\cite{Roshan2016}
\begin{equation}\label{eq10}
 G_{\alpha\beta}+\Lambda g_{\alpha\beta}=\kappa \big(\mathcal{F}^{\rho}_{\alpha}\mathcal{F}_{\rho\beta}-
 \frac{1}{4}g_{\alpha\beta}\mathcal{F}^{2}\big)+2\eta \big(\frac{1}{16}g_{\alpha\beta}(\mathcal{F}^{2})^{2}+
 2 \mathcal{F}^{\rho\sigma}\mathcal{F}^{\delta}_{\rho}\mathcal{F}_{\sigma\alpha}\mathcal{F}_{\delta\beta}-
 \frac{1}{4}g_{\alpha\beta}\mathcal{F}^{\gamma}_{\varsigma}\mathcal{F}_{\gamma\delta}\mathcal{F}^{\sigma\varsigma}\mathcal{F}^{\delta}_{\sigma} \big)\,,
\end{equation}
where $\mathcal{F}^{2}=\mathcal{F}_{\mu\nu}\mathcal{F}^{\mu\nu}$. So, in the high curvature regime i.e.
black hole, EMSG adds some modification terms to the matter field equations. Following this, the Maxwell
equations in the vacuum will be modified as follows
\begin{equation}\label{eq11}
\nabla_{\alpha}\mathcal{F}^{\alpha\beta}=\frac{\eta}{\kappa}\nabla_{\alpha}(4\mathcal{F}^{\beta}_{\rho}
\mathcal{F}^{\alpha\sigma}\mathcal{F}^{\rho}_{\sigma}-\mathcal{F}^{\alpha\beta}\mathcal{F}^{2})\,,
\end{equation}
\begin{equation}\label{eq12}
\nabla_{[\alpha}\mathcal{F}_{\beta\sigma]}=0\,,
\end{equation}
where $\nabla_{\alpha}$ is a covariant derivative with respect to $x^{\alpha}$. Equations (\ref{eq10}),(\ref{eq11}) and (\ref{eq12})
provide a complete set of coupled differential equations, which simultaneous solutions of them determine the
geometrical shape of the corresponding spacetime. Therefore, by considering a spherically symmetric line element as
\begin{equation}\label{eq13}
ds^{2}=g_{\alpha\beta}dx^{\alpha}dx^{\beta}=-b(r)dt^{2}+\frac{1}{b(r)}dr^{2}+r^{2}(d\theta^{2}+\sin^{2}\theta d\phi^{2})\,,
\end{equation}
and the only non-vanishing components of $\mathcal{F}_{\alpha\beta}$, that is, $\mathcal{F}_{tr}=-\mathcal{F}_{rt}=\mathcal{E}(r)$,
(where $\mathcal{E}$ is the electric field), the exact solution of the field equations will specify the $b(r)$ as follows~\cite{Roshan2016}
\begin{equation}\label{eq14}
b(r)=1-\frac{\kappa M}{4\pi r}-\frac{\Lambda r^{2}}{3}-\frac{\kappa}{2r}\int \big(\mathcal{E}^{2}(r)+
\frac{3\eta}{\kappa}\mathcal{E}^{4}(r) \big) r^{2}dr\,,
\end{equation}
where $M$ is the mass of the charged black hole. It is easy to see that the Reissner-Nordstrom metric is recovered
when $\frac{\eta}{\kappa}$ is small. For $\frac{\eta}{\kappa}\ll1$, the $b(r)$ approximately takes the following form
\begin{equation}\label{eq15}
  b(r)=1-\frac{\Lambda r^{2}}{3}-\kappa \big(\frac{M}{4\pi r}-\frac{q^{2}}{2r^{2}}+\frac{q^{4}}{10r^{6}}\frac{\eta}{\kappa}\big)\,.
\end{equation}
By checking the above metric, we find that $r=0$ is a singularity for this spacetime metric. Nevertheless, it is an exterior
solution for a charged black hole spacetime and therefore, the $r=0$ singularity will be covered by the event horizon, $r_{eh}$.
After a brief introduction of the EMSG charged black hole, We note that in such a nonlinear framework,
photons propagate along geodesics that are not null in the Minkowski spacetime. In fact these geodesics are null in an effective geometry~\cite{Novello2000,Costa2009}. In order to study optical features of this black hole we derive the effective metric for this spacetime.

\section{THE EFFECTIVE GEOMETRY FOR EMSG charged black hole}\label{se9}

Nonlinear electrodynamics plays a substantial role in diverse physics fields. In quantum field theory, vacuum polarization inherently brings about nonlinear adjustments to Maxwell’s electrodynamics, as outlined by the Euler-Heisenberg Lagrangian. In substances like specific insulators and gems, the complex connections between molecules and external electromagnetic fields can be accurately represented by a nonlinear theory, frequently seen at high light intensities, such as those generated by pulsed lasers.

The discovery that high-energy disturbances in nonlinear electromagnetic theory propagate along geodesics in an effective spacetime, instead of null geodesics in the background geometry, has been consistently validated in the literature. For the scenario when $( L = L(\mathcal{F}) )$, the motion equation is expressed as:

\begin{equation}\label{a}
  \partial_{\mu}(\frac{\partial L}{\partial \mathcal{F} ^{\mu\nu}})=0
\end{equation}

This expression illustrates how the fields change under the impact of nonlinearities in the theory. By perturbing the equation \ref{a} around a fixed background solution and taking the eikonal limit \cite{Novello2000}, the effective metric takes the following form

\begin{equation}\label{b}
  \mathfrak{g}^{\mu\nu}=4L_{\mathcal{F}\mathcal{F}}\mathcal{F}^{\mu}_{\alpha}\mathcal{F}^{\alpha\nu}-L_{\mathcal{F}}g^{\mu\nu}
\end{equation}

where $L_{\mathcal{F}}=\frac{d L}{d \mathcal{F}}$  and  $L_{\mathcal{F}\mathcal{F}}=\frac{d^{2} L}{d \mathcal{F}^{2}}$. By using the equation \ref{eq14}, and considering the $L(\mathcal{F})=-\frac{\eta}{2\kappa}\mathbf{T}^{2}$ we derive the effective metric for EMSG charged black hole

\begin{equation}\label{c}
  ds^{2}_{eff}= A [- b(r) dt^{2}+ b^{-1}(r) dr^{2}+h (d\theta^{2}+\sin^{2}(\theta)d\phi^{2})]\,
\end{equation}
where the $A , h $ define as follows

\begin{equation}\label{d}
  A=(-6\frac{\eta}{\kappa} \big(\frac{q}{r^{2}}-2\frac{q^{3}}{r^{6}}\frac{\eta}{\kappa})\big)^{-1}\big(\frac{4}{3}(\frac{q}{r^{2}}-2\frac{q^{3}}{r^{6}}\frac{\eta}{\kappa})+1\big)^{-1}, \quad\quad   h=\big(\frac{4}{3}(\frac{q}{r^{2}}-2\frac{q^{3}}{r^{6}}\frac{\eta}{\kappa})+1\big)r^{2}\,.
\end{equation}

In the following by considering the effective metric Eq. (\ref{c}) we investigate the shadow behavior of EMSG charged black hole.

\section{PHOTON MOTION IN EMSG CHARGED BLACK HOLE}\label{se3}

Consider a photon in the background of EMSG charged black hole. We note that in such a nonlinear framework,
photons propagate along geodesics that are not null in the Minkowski spacetime. In fact these geodesics are null in the effective geometry which is characterized by metric function as Eq. (\ref{c}).
In this effective geometry photon follows a null geodesics through the following Lagrangian
\begin{equation}\label{eql}
\mathfrak{L}=\frac{1}{2}A\Big(-b(r)\dot{t}^{2}+\frac{1}{b(r)}\dot{r}^{2}+h(\dot{\theta}^{2}+\sin^{2}\theta\dot{\phi}^{2})\Big)\,,
\end{equation}
where an over dot denotes the derivative with respect to the affine parameter $\sigma$.
The Components of canonically conjugate energy-momentum relevant to the Lagrangian (\ref{eql}) is determined as follows
 \begin{equation}
\mathbf{P}_{t}=\frac{\partial \mathfrak{L}}{\partial \dot{t}}=A b(r)\dot{t}=E,
\end{equation}
\begin{equation}
 \mathbf{P}_{r}= \frac{\partial \mathfrak{L}}{\partial \dot{r}}=A b^{-1}(r)\dot{r},
\end{equation}
\begin{equation}
\mathbf{P}_{\phi}=\frac{\partial \mathfrak{L}}{\partial \dot{\phi}}=A h \sin^{2}\theta \dot{\phi}=L,
\end{equation}
and
\begin{equation}
 \mathbf{P}_{\theta}= \frac{\partial \mathfrak{L}}{\partial \dot{\theta}}= A h\dot{\theta},
\end{equation}
in which $E$ and $L$ stand for the energy and angular momentum of the photon, respectively.
To analyze the complete geodesic equations for photon motion we follow the Hamilton-Jacobi method and Carter
constant separable approach~\cite{Carter1968}. The most general form of the Hamilton-Jacobi equation is expressed as
\begin{equation}
 \frac{\partial S}{\partial \tau}=-H=-\frac{1}{2}\mathfrak{g}^{\alpha\beta}\frac{\partial S}{\partial x^{\alpha}} \frac{\partial S}{\partial x^{\beta}},
\end{equation}
where $S$ is the Jacobian action for the photon. Using the metric (\ref{eq15}), we obtain
\begin{equation}\label{eq20}
 -2\frac{\partial S}{\partial \tau}= (-Ab)^{-1}(r)(\frac{\partial S_{t}}{\partial t})^{2}+
 A^{-1} b(r)(\frac{\partial S_{r}}{\partial r})^{2}+A^{-1} h^{-1}\big[(\frac{\partial S_{\theta}}{\partial \theta})^{2}+
  \frac{1}{\sin^{2}(\theta)}(\frac{\partial S_{\phi}}{\partial \phi})^{2}\big]\,.
\end{equation}

Now, adopting the separable solution for the Jacobi action allows us to represent the action as follows
\begin{equation}\label{EQW}
S=-Et+L\phi+S_{r}(r)+S_{\theta}(\theta)\,.
\end{equation}
The energy $E$ and angular momentum $L$ are two conserved quantities associated with
temporal translation and rotation around the axes of symmetry, respectively. Along with these
obvious constants of motion, Carter \cite{Carter1968} demonstrated the existence of an
additional conserved quantity by separating the Hamilton-Jacobi equation, known as the Carter constant.
By inserting the Carter constant and
the expression~(\ref{EQW}) in equation~(\ref{eq20}), we find
\begin{equation}
 0=\big[- b^{-1}(r)(E)^{2}+b(r)(\frac{\partial S_{r}}{\partial r})^{2}+ h^{-1}(\mathcal{K}+L^{2})\big]-
  \big[h^{-1}\big((\frac{\partial S_{\theta}}{\partial \theta})^{2}-\mathcal{K}+L^{2}\cot^{2}(\theta)\big)\big]\,.
\end{equation}
After some calculation, the above equation will be separated into the following two equations;
 \begin{equation}\label{eq21}
 h^{2}b^{2}(r)\left(\frac{\partial S_{r}}{\partial r}\right)^{2}=\left[E^{2}-\frac{b(r)}{h}(L^{2}+\mathcal{K})\right]h^{2}\,,
 \end{equation}
and
\begin{equation}\label{eq22}
 (\frac{\partial S_{\theta}}{\partial \theta})^{2}=\mathcal{K}-L^{2}\cot^{2}(\theta)\,.
\end{equation}
Finally by employing equations (\ref{eq21}) and (\ref{eq22}) along with the components of
the canonically conjugate momentum, we derive the complete set of equations for the photon motion in the background of EMSG charged black hole as
\begin{equation}
   \dot{t}=\frac{E}{A b(r)}\,,
 \end{equation}
  \begin{equation}\label{eq23}
 h \dot{r}=\pm\sqrt{\mathcal{R}(r)}\,,
 \end{equation}
 \begin{equation}
 \dot{\phi}=\frac{L}{r^{2}\sin^{2}(\theta)}\,,
 \end{equation}
 \begin{equation}
  \dot{\theta}=\pm\sqrt{\Theta(\theta)}\,,
\end{equation}
where the outgoing and ingoing radial directions of the motion of the photon
are distinguished by the ``+"  and  ``-" signs respectively. For the null geodesics, the involved quantities $\mathcal{R}(r)$ and $\Theta(\theta)$ are defined as follows
\begin{equation}
   \mathcal{R}(r)=E^{2} \frac{h^{4}}{A^{2}}-b(r)\frac{h}{A^{2}}(L^{2}+\mathcal{K})\,,
 \end{equation}
 \begin{equation}
 \Theta(\theta)=  \mathcal{K}-L^{2}\cot^{2}(\theta)\,.
 \end{equation}
Depending on the energy and angular momentum of the photon approaching a black hole, there
are three different destinations for the photon: it may be scattered, it may be captured, or it may end up in unstable orbits around the black hole.

The mathematical tool to study the photon motion more precisely is the effective potential, which can be read by the radial null geodesics equation~(\ref{eq23})
\begin{equation}
   (\dot{r})^{2}+V^{(eff)}=0\,,
 \end{equation}
in which the effective potential of the photon is defined as
\begin{equation}
  V^{(eff)}=-\frac{E^{2}}{A^{2}}+\frac{b(r)}{A^{2}h}(\mathcal{K}+L^{2})\,.
\end{equation}
\begin{figure}[H]
\centering
\includegraphics[width=0.7\textwidth]{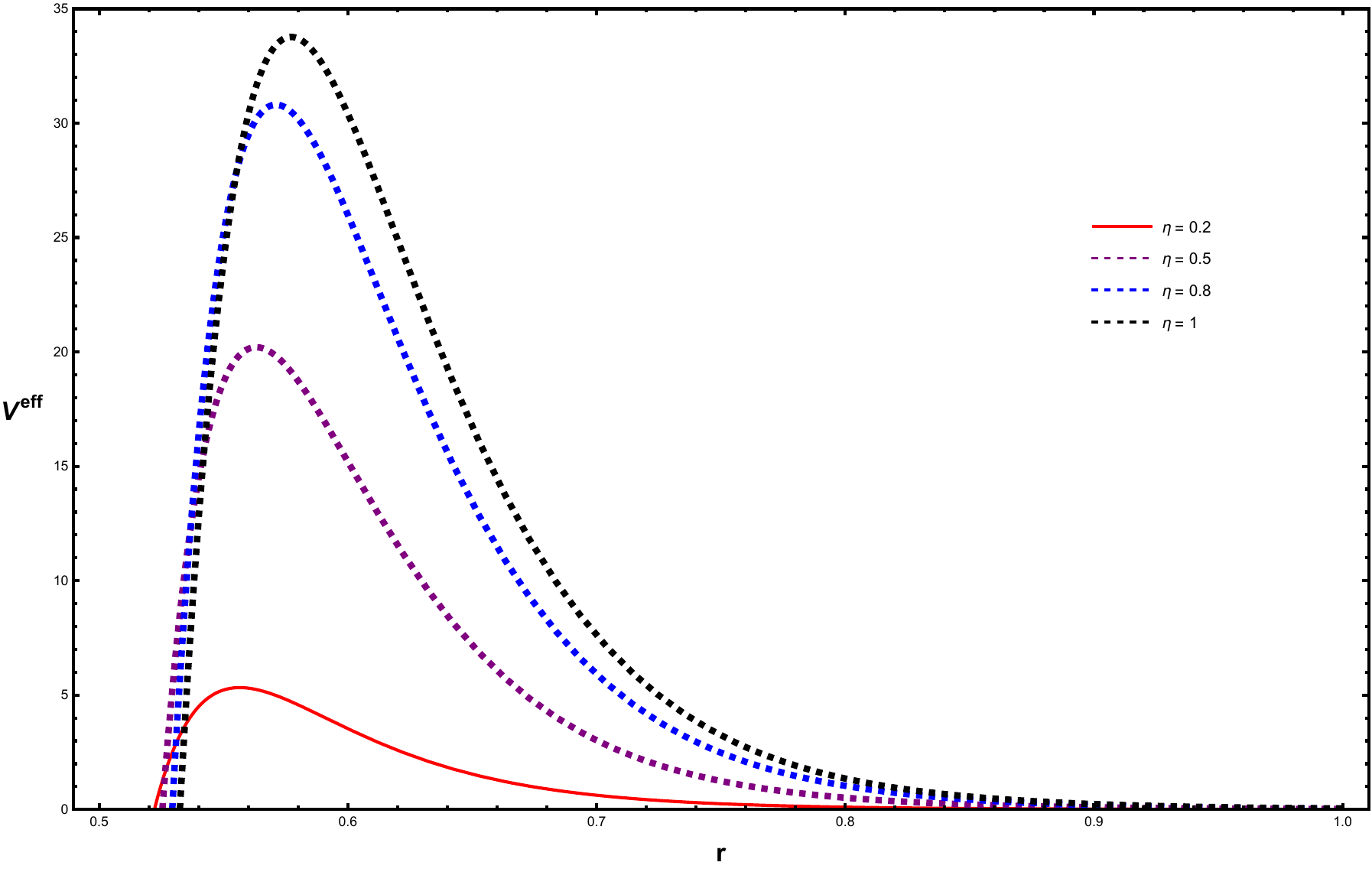}
\caption{\label{Fig1}\small{\emph{ The illustration of $V^{(eff)}$ as a function of $r$ for $Q=0.5$,\, $\Lambda=0.01$ and different values of $\eta$.}}}
\end{figure}
Fig.~\ref{Fig1} demonstrates the behavior of the effective potential for the photon moving in
the EMSG charged black hole spacetime as a function of the radial coordinate $r$ for different values of the parameter $\eta$.
From this figure we see that the maximum value of the effective potential increase by
increment the values of $\eta$ and in the limit of $r \rightarrow \infty$, the effective potential approaches a constant value.
In fact, at large radial distances the effect of $\eta$ tends to zero and GR's result is recovered.\\
Among all possible motions of the photons, that is, capturing, scattering and unstable (circular/spherical) orbits,
only unstable orbits assume a part in the formation of the black hole shadow. The unstable circular orbits of
the photons happen at a distance called the \emph{photon sphere radius} $r_{ph}$ which corresponds
with the maximum value of the effective potential and can be determined by the following equations
\begin{equation}\label{eq24}
  V^{(eff)}|_{r_{ph}}=0=\frac{dV^{(eff)}}{dr}|_{r_{ph}}, \quad\quad     \mathcal{R}(r)|_{r_{ph}}=\frac{dR(r)}{dr}|_{r_{ph}}=0\,.
\end{equation}
It is worth mentioning that since the strong gravitational field in the vicinity of the black hole
causes the gravitational lensing, the boundary of the black hole shadow does not occur at the unstable photon orbits,
but at the apparent shape of the unstable photon orbits. Subsequently,
the radius $r_{ph}$ of the photon sphere for EMSG charged black hole is the smallest value of the roots of the following equation
\begin{equation}\label{epr}
  b' A h-2b A' h- b A h'+2b h A'=0\,.
\end{equation}
In which prime denotes the derivative respect to $r$.  This smallest root can be obtained numerically as can be seen in what follows.

\section{GEOMETRICAL SHAPE OF THE SHADOW OF EMSG CHARGED BLACK HOLE}\label{se4}

\begin{figure}[H]
\centering
\includegraphics[width=0.7\textwidth]{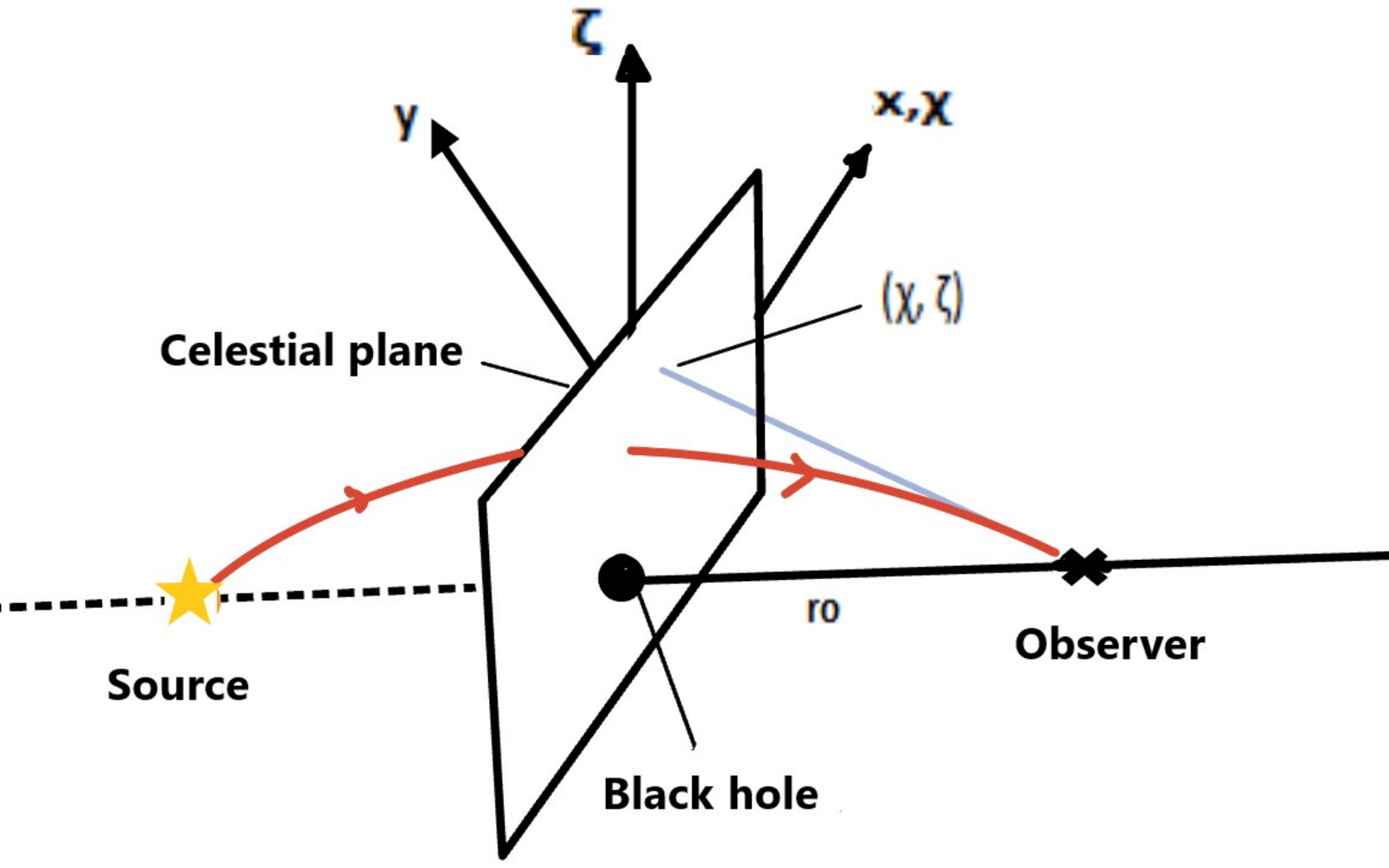}
\caption{\label{Fig2}\small{\emph{ The schematic diagram of celestial coordinates from observer sight located far from the black hole in the position $(r_{o},\frac{\pi}{2})$, where $r_{o}$ determines the distance from the black hole and $\frac{\pi}{2}$ is the angular coordinate for the observer.  The coordinates $(\chi,\zeta)$ are the apparent distances of the image from the symmetry axis and from its projection on the equatorial plane, respectively. }}}
\end{figure}

The apparent shape of a black hole is completely distinguishable by the constraint on its shadow.
The photons moving in unstable orbits around a black hole form shadow which appears as two
dimensional perfect dark disc for a spherically symmetric black hole. The size and shape of
a shadow can be determined by geometrical optics considerations. For this purpose, we introduce two impact
parameters $a$ and $b$ to characterize a photon near the black hole. These impact parameters are defined as follows
\begin{equation}\label{eq26}
a=\frac{L}{E}\,, \quad\quad    b=\frac{\mathcal{K}}{E^{2}}\,.
\end{equation}
Therefore, we can rewrite $V^{(eff)}$ and $ \mathcal{R}(r)$ as functions of these impact parameters as follows
\begin{equation}\label{eq27}
V^{(eff)}=\Big\{\frac{b(r)}{h}(a^{2} +b)-1\Big\}\frac{E^{2}}{A^{2}}\,, \quad\quad   \mathcal{R}(r)=\Big\{-(a^{2} +b)h b(r)+h^{2}\Big\}\frac{E^{2}}{A^{2}}\,.
\end{equation}
Having $V^{(eff)}$ and $ \mathcal{R}(r)$, through equation (\ref{eq24}) we obtain the following equation governing the parameters $a$ and $b$
\begin{equation}
a^{2} +b=\frac{2 h h' A^{2}-2 A A' h^{2}}{h' b A^{2}+h b' A^{2}-h b 2 A A'}\,,
\end{equation}
where the radius of the photon sphere is obtained from equation (\ref{epr}).
In order to visualize the geometrical shape of the shadow on the observer’s frame, it is convenient to employ
the celestial coordinates $ \chi$ and $\zeta$ which are defined as~\cite{Vazquez2004}
\begin{equation}
\chi=\lim_{r_{o}\rightarrow \infty}(\frac{r^{2}_{o}p^{\phi}}{p^{t}})=\frac{-a}{\sin\theta}\,,
\end{equation}
and
\begin{equation}
\zeta=\lim_{r_{o}\rightarrow \infty}(\frac{r^{2}_{o}p^{\theta}}{p^{t}})=\pm\sqrt{b-a^{2}\cot(\theta)}\,,
\end{equation}
in which $ r_{o}$ is the distance between the black hole and a far distant observer and $[p^{t},p^{\theta},p^{\phi}]$
are the tetrad components of the momentum. Fig.~\ref{Fig2} shows a schematic of the celestial coordinates used in this setup.

One can see where $\theta=\frac{\pi}{2}$ (equatorial plane), then
$\chi=-a$ and $\zeta=\pm\sqrt{b}$. Accordingly, we reach at the following result
\begin{equation}\label{eq30}
  R_{s}^{2}=a^{2} +b=\chi^{2}+\zeta^{2}\,,
\end{equation}
where $R_{s}$ is the shadow radius in celestial coordinates; the shadow of which appears as a circle.
In table \ref{t1} we collect some computed numerical results for the event horizon ($r_{h}$), cosmological horizon ($r_{c}$),
photon sphere radius $r_{ph}$ and shadow radius $R_{s}$ for an EMSG charged black hole.

\begin{figure}[H]%
    \centering
    \subfloat[\centering Q=0.1]{{\includegraphics[width=8cm]{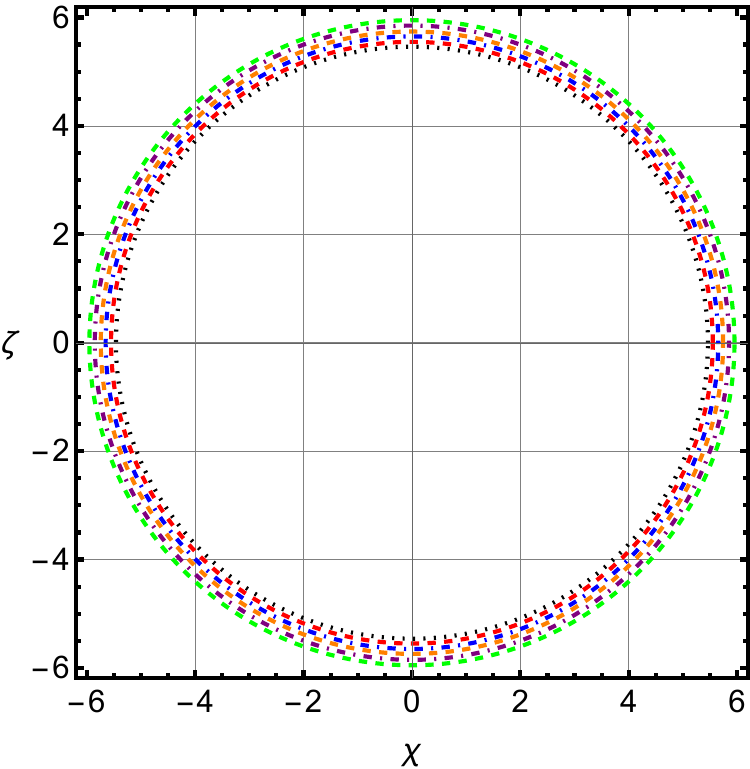} }}%
    \qquad
    \subfloat[\centering $\eta$=0.01]{{\includegraphics[width=8cm]{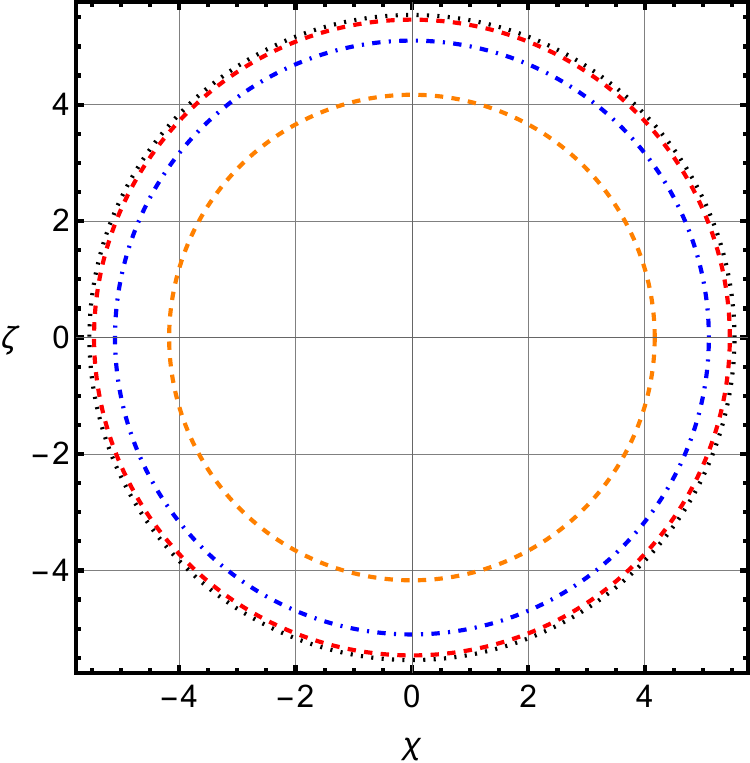} }}%
    \caption{Contour plot of geometrical shape of EMSG charged black hole in celestial coordinate with $\Lambda = 0.01$}%
    \label{vb}%
\end{figure}

Based on the numerical results provided in table~\ref{t1}, we draw a contour plot of the equation (\ref{eq30}). Fig.~\ref{vb}
demonstrates the geometrical shapes of the shadow of the EMSG charged black hole for $\Lambda=0.01$
in celestial coordinates. Fig.~(\ref{vb}a) shows the shadow of the EMSG charged black
hole for different values of the parameter $\eta$. This figure shows that the shadow size for different values of the parameter $\eta$
approach each other. Fig.~(\ref{vb}b) is an illustration of the shadow of the EMSG charged black
hole for different values of parameter $Q$. From Fig.~(\ref{vb}a) we see that by enhancing the effect of the electric charge, the shadow size decreases accordingly.
About the role of the cosmological constant on the shadow, generally the shadow angle decreases by increasing the value of the cosmological constant so that the shadow angle decreases faster for the larger values of the cosmological constant in this setup. To see more details about the influence of the cosmological constant on the shadow of a black hole we refer to~\cite{Perlick2018,Firouzjaee2019}. As has been shown in Ref.~\cite{Perlick2018}, the angular radius of the shadow of a Schwarzschild-de Sitter black hole shrinks to a non-zero
finite value if the comoving observer approaches infinity. So, the limit of infinity in Eqs. (41) and (42) is defined for an observer who is comoving with the cosmic expansion.

\begin{table}
\begin{center}\label{t1}
\caption{\label{t1} The computed numerical values for $r_{h}$, $r_{ph}$, $r_{c}$ and $R_{s}$ of EMSG charged Black hole for different values of the parameters $Q$ and $\eta$ }
\begin{tabular}{ |p{2cm}||p{1cm}|p{1cm}|p{1cm}| p{1cm}| p{1cm}| p{1cm}||p{2cm}||p{1cm}|p{1cm}|p{1cm}|p{1cm}| }
 \hline

  \multicolumn{7}{||c||}{Q = 0.1  }&
  \multicolumn{5}{c||}{$\eta = 0.01$  }
  \\[2ex]

 \hline\hline
 $\eta$ & $0.01$ & $0.02$ & $0.03$ & $0.04$& $0.05$& $0.06$ & $Q$ & 0.01&0.1 &0.2 &0.3 \\ [1ex]
 \hline\hline\hline
 $r_{h}$    & 1.98    & 2 &  2.03 &2.05& 2.07&2.1 & $r_{h}$& 2.05&1.98& 1.75& 0.473\\[1ex]
 \hline
 $r_{ph}$  &   2.94  & 2.98   &  3.01 &3.04& 3.07& 3.11 &$r_{ph}$& 3.03& 2.94& 2.65& 1.321\\[1ex]
 \hline
 $r_{c}$  &16.1 & 16&  15.9& 15.8&15.7&15.6 & $r_{c}$& 16.1&16.1&16.1& 16.2\\[1ex]
 \hline
$R_{s}$    &5.46 &5.55& 5.65 &5.74& 5.85&5.95& $R_{s}$ &5.54&5.46&5.10&4.17\\[1ex]
 \hline\hline\hline
\end{tabular}
\end{center}
\end{table}

\section{CONSTRAINTS FROM EHT OBSERVATION OF Sgr A$^{\ast}$} \label{se5}

\begin{figure}%
    \centering
    \subfloat[\centering Q=0.1]{{\includegraphics[width=8cm]{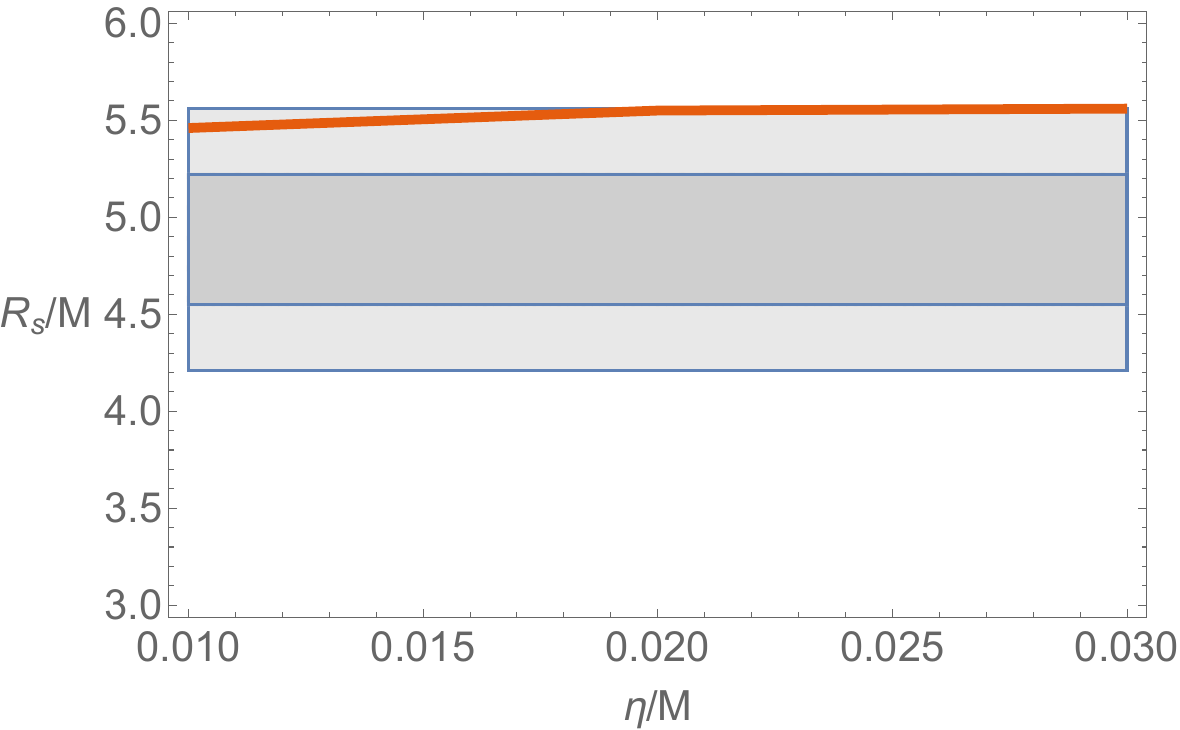} }}%
    \qquad
    \subfloat[\centering $\eta$=0.01]{{\includegraphics[width=8cm]{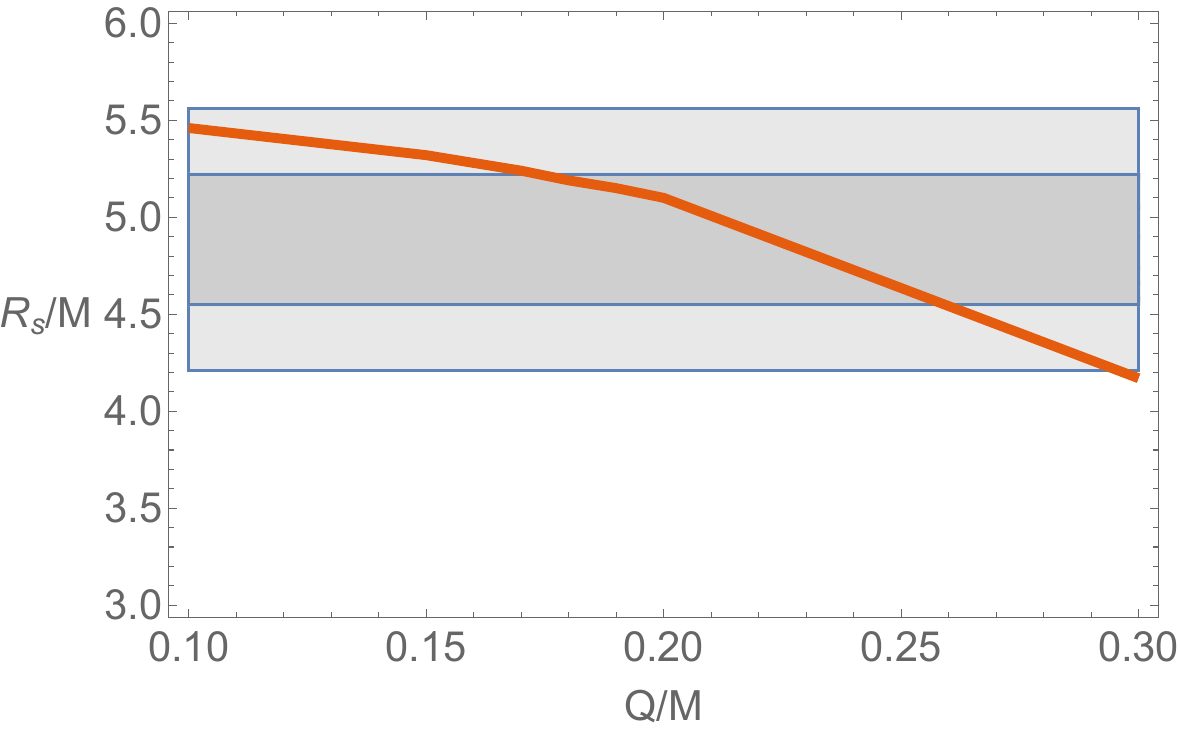} }}%
    \caption{The shadow radius of an EMSG charged black hole in units of the black hole mass $M$.}%
    \label{fig4}%
\end{figure}

In this section we compare our results of the shadow size of the EMSG charged black hole with the recent
image of the supermassive black hole Sgr A* released by the EHT Collaboration. Indeed, we investigate the possible
constraints on the Energy-Momentum Squared Gravity by utilizing the EHT provided data.
The captured image by EHT collaboration delivered through the Very Long Baseline Interferometry (VLBI), shows
a bright emission ring surrounding a central dark spot. It is interesting to note that very
recently EHT collaboration has revealed strong magnetic fields spiraling at the edge of Milky Way's central black hole,
Sgr A*~\cite{EHT2024}, the observation of which in some sense supports the idea to study charged version of the EMSG black holes in the present manuscript.
We can use the size of the bright ring to estimate the size of the black hole shadow under two
circumstances: $i$) A bright source of light exists that its light strongly bends near the horizon,
and $ii$) The surrounding emission region is geometrically thick and optically
thin at the VLBI network wavelength \cite{Vagnozzi2022moj}.\\
For this purpose, first we need the mass-to-distance ratio of Sgr A*.
These two quantities have been measured by two sets of instruments/teams, ``Keck" and ``VLTI".
They found their results through the stellar cluster dynamics model by S$0-2$ star which
is shown in table~\ref{t2} at the $68$$\%$  level of confidence~\cite{Doetal2019},\cite{Abuter2020}.
\begin{table}[H]
\begin{center}\label{t2}
\caption{\label{t2} Mass and distance of Sgr A* estimated via Keck and VLTI instruments.}
\begin{tabular}{ |p{2cm}||p{4cm}|p{4cm}| }
 \hline
 $\textrm{Survey}$ & $M (\times 10^{6} M_{\odot})$ & $D (kpc)$  \\ [1ex]
 \hline\hline\hline
 $Keck$    & 3.951 ± 0.047    & 7.953 ± 0.050 ± 0.032  \\[1ex]
 \hline
 $VLTI$  &   4.297 ± 0.012 ± 0.040  &8.277 ± 0.009 ± 0.033   \\[1ex]
 \hline
\end{tabular}
\end{center}
\end{table}
The second requirement is the calibration factor to link the size of the bright emission ring with the size of
the corresponding shadow which tells us how reliable it is to use the size of the bright ring of
emission as an indicator of the shadow size. Several sources of uncertainties e.g. formal measurement
uncertainties, fitting/model uncertainties, theoretical uncertainties pertaining to the emissivity of
the plasma~\cite{Vagnozzi2022moj} (see the original discussion in Sec. 3 of Ref.~\cite{Akiama2022}),
produce the calibration factor. The EHT conducts
the calibration factor with uncertainties in Sgr A*’s mass-to-distance ratio,
and the angular diameter of Sgr A*’s bright ring of emission by carefully examining different sources of
uncertainties \cite{Akiama2022}. The fractional deviation between the deduced
shadow radius $r_{sh}$ and the shadow radius of a Schwarzschild black hole of angular size $\theta$ ($\theta_{sh}=3\sqrt{3}\theta_{g}$),
$r_{sh,Schwarzschild} = 3\sqrt{3}M$ reads $\delta$ which are collected by EHT. In practice $\delta$ is given as follows~\cite{Vagnozzi2022moj}
\begin{equation}\label{delta}
  \delta=\frac{r_{sh}}{r_{sh,Schwarzschild}}-1=\frac{r_{sh}}{3\sqrt{3}M}-1\,.
\end{equation}
Based on Keck and VLTI measurements of the mass-to-distance ratio, $\delta$ will be estimated as $0.04^{+0.09}_{-0.10} $
for Keck and $0.08^{+0.09}_{-0.09} $ for VLTI. For simplicity, one can
take the average of the Keck and VLTI-based estimates of $\delta$ that leads to the following approximate value
\begin{equation}
  \delta \simeq - 0.060 \pm 0.065\,.
\end{equation}
Under the assumption of Gaussianity, this gives easily the following $1\sigma$ and $2\sigma$ ranges for $\delta$
\begin{equation}
\begin{split}
-0.125 & \leq \delta \leq 0.005  \quad (1\sigma)\,, \\
-0.19 & \leq\delta \leq 0.07      \quad   (2\sigma)\,.
\end{split}
\end{equation}
According to the Keck and VLTI  estimations, we sense that Sgr A*’s shadow is slightly smaller than
the expected $3\sqrt{3}M$ for a Schwarzschild black hole with the same mass, with a  $\leq 68$$\%$ level of confidence
preference for $\delta < 0$. By assuming Gaussian uncertainties, one can consider equation (\ref{delta}) and
inverting $r_{sh}/M = 3\sqrt{3}(\delta + 1)$ to find the following $1\sigma$ and $2\sigma$ limits on $r_{sh}/M$:
\begin{equation}\label{bou}
\begin{split}
4.55 & \leq r_{sh}/M \leq 5.22  \quad (1\sigma)\,, \\
4.21 & \leq r_{sh}/M \leq 5.56     \quad    (2\sigma)\,.
\end{split}
\end{equation}
Now we utilize the limits in equations (\ref{bou}) to compare the shadow of the EMSG charged black hole with Sgr A*'s shadow. Fig.~\ref{fig4} illustrates
the behavior of the shadow radius of the EMSG charged black hole in comparison with the EHT’s shadow
size of Sgr A* within $1\sigma$ and $2\sigma$ levels of confidence. The gray and light gray areas are
compatible with the EHT image of Sgr A* at $1\sigma$ and $2\sigma$ levels of confidence respectively, after taking
into account the Keck and VLTI average mass-to-distance ratio for Sgr A*. More than $2\sigma$
uncertainties are excluded regions by the same observations. In Fig.~\ref{fig4}a, we plotted the
shadow radius of EMSG charged black hole as a function of $\eta$ for $Q = 0.1$ and $\Lambda = 0.01$.
As it is clear, for $\eta \leq 0.02$ the EMSG charged black hole shadow radius is compatible
with Sgr A* in $2\sigma$ level of confidence. According to Fig.~\ref{fig4}b, for $\eta = 0.01$ and $\Lambda = 0.01$
in the interval $0.01 \leq Q \leq 0.16$ the shadow radius of the EMSG charged black hole stays
within the $2\sigma$ limit. In the interval $0.16 \leq Q \leq 0.26 $, the shadow radius of the EMSG charged black
hole satisfies the $1\sigma$ confidence level of the EHT observations. Therefore, the Sgr A* can be an EMSG charged
black hole in an accelerating universe ($\Lambda = 0.01$) in the units of $\eta = 0.01$, $Q = 0.1$.

\section{ENERGY EMISSION RATE}\label{se7}

Hawking radiation is a phenomenon that allows black holes to emit radiation
through the quantum fluctuation process.
In fact inside the black hole and near the horizon, quantum fluctuations create and annihilate
many pairs of particles-antiparticles. In this regard, particles with positive energy can escape from the black
hole through a quantum tunneling process, which can lead the black hole to radiate and evaporate gradually.
The associated energy emission rate can be expressed as follows~\cite{wei2013}
\begin{equation}\label{eqe}
 \frac{d^{2}E(\omega)}{d\omega dt}=\frac{2\pi^{2}\sigma_{lim}}{\exp(\omega/T)-1}\omega^{3}\,,
\end{equation}
in which $\omega$ is the emission frequency and $T$ is the Hawking temperature of the event horizon which we obtain as
\begin{equation}
 T = \frac{(5 M r_{h}^{5}+4 \pi(3 q^{4}-5 q^{2}r_{h}^{4}))\kappa}{80 \pi^{2}r_{h}^{7}}-\frac{2 r_{h} \Lambda}{3}\,.
\end{equation}
In the high energy limit, the absorption cross section oscillates around a limiting constant value $\sigma_{lim}$
which is approximately equal to the photon sphere or black hole shadow. This value reads as (see Refs.~\cite{Decanini2011,Decanini2011a,Mashhoon1973})
\begin{equation}
  \sigma_{lim}\approx \pi R_{s}^{2}\,,
\end{equation}
where $R_{s}^{2}$ is the black hole shadow radius (\ref{eq30}).\\
\begin{figure}%
    \centering
    \subfloat[\centering Q=0.1]{{\includegraphics[width=8cm]{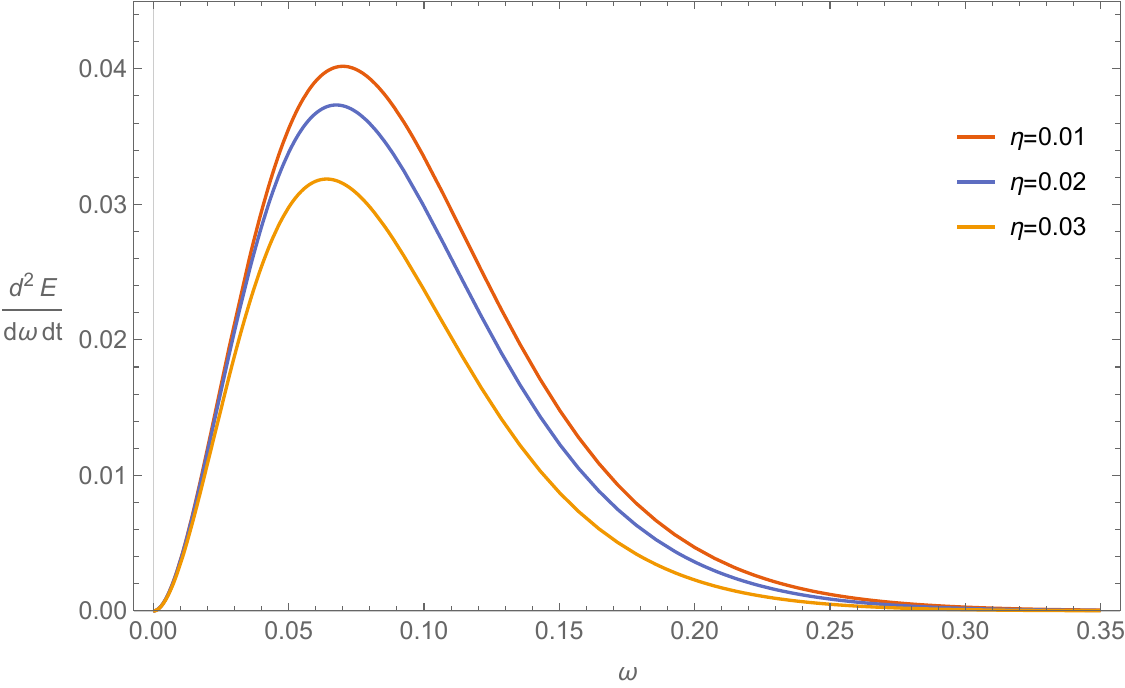} }}%
    \qquad
    \subfloat[\centering $\eta$=0.01]{{\includegraphics[width=8cm]{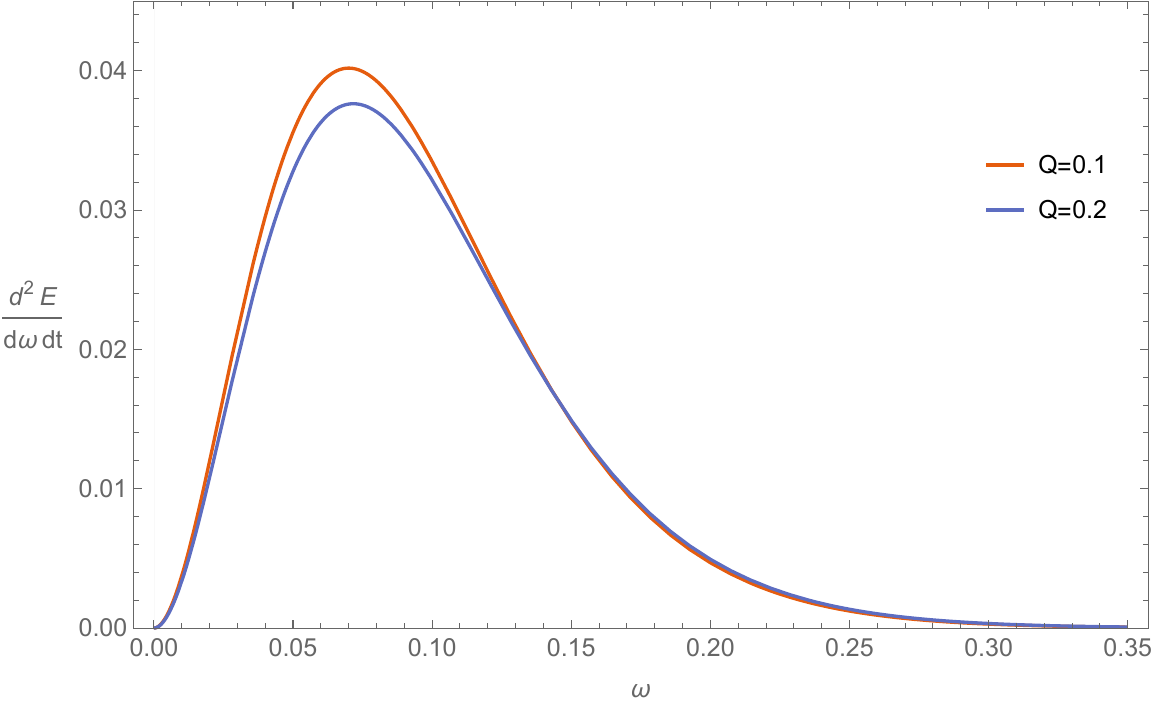} }}%
    \caption{The graph of energy emission rate as function of $\omega$ for EMSG charged black hole}%
    \label{fig3}%
\end{figure}
Fig.~(\ref{fig3})is the illustration of the energy emission rate form a EMSG charged black hole as a function of $\omega$.
In Fig.~(\ref{fig3}a) we plotted the energy emission rate of the EMSG charged black hole for different values of the parameter $\eta$ by
setting $Q=0.1$. We find out that increment of $\eta$ decreases the energy emission rate for the EMSG charged black hole.
So, the evaporation of the EMSG charged black hole decelerates by growth of the parameter $\eta$. Fig.~(\ref{fig3}b) shows the energy emission rate from a
EMSG charged black hole for different values of the charge parameter $Q$, when $\eta$ is fixed. We see that the
energy emission rate for the EMSG charged black hole will decrease, which means that the
EMSG charged black hole evaporates faster for the smaller values of the electric charge.

\section{SUMMERY AND CONCLUSION }\label{se8}

In this work we have investigated the shadow cast formed by an EMSG charged black hole in confrontation with EHT data of Sgr A*.
We started by focusing on the motion and trajectories of photons in the background of an EMSG charged black hole. In non-linear electrodynamics, photons move along geodesics that are null in the effective metric. Therefore, we first obtained the effective metric for an EMSG charged black hole.
For this purpose, we utilized the Hamilton-Jacobi method and Carter constant
to find complete equations of motion for photons' trajectories, and derivation of the corresponding
effective potential. Then, by introducing appropriate celestial coordinates and by deriving the effective potential
to calculate the radius of the possible circular orbits, we demonstrated the shadow shape of
EMSG charged black hole by some constructive contour plots. Indeed, we plotted the shadow shape of the EMSG charged black hole in celestial
coordinate for positive values of the cosmological constant to fulfill the late time cosmic speed up.
We have shown that for a given electric charge, the shadow size of the EMSG charged black hole
varies effectively for different values of the parameter $\eta$. We also found out that for a fixed value of $\eta$,
the shadow size increases by increasing the electric charge. Then we used EHT data to evaluate and justify
our results of the shadow shape in confrontation with Sgr A* data. As a result and within the parameter space of the present model, the Sgr A* can be an EMSG charged
black hole in an accelerating universe (for instance, with $\Lambda = 0.01$, $\eta = 0.01$ and $Q = 0.1$).
We have constrained the parameter $\eta$ and the electric charge by comparing the shadow size reported by EHT for Sgr A* with
the shadow radius of the EMSG charged black hole calculated in this paper. We found that for a fixed value of the electric charged, the shadow radius of
the EMSG charged black hole for $0.01\leq\eta\leq0.02$ lies within the $2\sigma$ confidence level of the Sgr A* shadow radius.
We also noticed that for $\eta= 0.01$ , the range for the electric charge of EMSG charged black hole to be
compatible with the EHT data of Sgr A* has to be $0.01\leq Q\leq0.3$. Furthermore, we explored the
energy emission rate for the EMSG charged black hole in this setup. We observed that for smaller
amounts of the electric charge and the parameter $\eta$, the EMSG charged black hole evaporates
faster. As EHT collaboration has reported very recently, strong magnetic fields are spiraling at the edge of Milky Way's central black hole,
Sgr A*~\cite{EHT2024}. This novel observation highlights the role of magnetic field in black hole spacetime which in some sense supports the idea to study charged version of the EMSG black holes in this study.

\textbf{Acknowlegment}

we appreciate Dr sara saghafi for fruitful discussion.

\end{document}